\documentclass[twocolumn,showpacs,preprintnumbers,amsmath,amssymb]{revtex4}

\usepackage{graphicx}
\usepackage{dcolumn}
\usepackage{bm}
\usepackage{epsfig}
\usepackage{color}

\begin{document}


\title{Irreversible Eshelby description of aging in glasses}

\author{U. Buchenau}
 \email{buchenau-juelich@t-online.de}
\affiliation{%
Forschungszentrum J\"ulich GmbH, J\"ulich Centre for Neutron Science (JCNS-1) and Institute for Complex Systems (ICS-1),  52425 J\"ulich, GERMANY
}%

\date{April 28, 2019}

\begin{abstract}
The recent description of the cooling through the glass transition in terms of irreversible structural Eshelby rearrangements with a single average fictive temperature is extended to a distribution of fictive temperatures around the average one. The extension has only little influence on the cooling scans, but turns out to be necessary to understand the heating back to equilibrium.
\end{abstract}

\maketitle

Glasses are frozen liquids, frozen in the liquid structure as the liquid falls out of the thermal equilibrium at the glass transition temperature $T_g$ \cite{ediger,angell,schick1,cavagna,schmelzer}. The glass transition is a kinetic transition, but resembles a second order phase transition due to the very strong temperature dependence of the viscosity. The viscosity even extrapolates to infinity at the Vogel-Fulcher temperature $T_{VF}$ lying not too far below $T_g$.

Describing the aging behavior of the frozen glass below $T_g$ requires the introduction of a fictive temperature \cite{tool,nara,greg,hodge} $T_f$, which is higher than the phonon temperature $T$ and characterizes the frozen state of the structural degrees of freedom.

The two most frequently employed phenomenological approaches for calculating the time development of the fictive temperature (or the corresponding deviation from the equilibrium density) in polymers and simple liquids are the Tool-Narayanaswamy-Moynihan (TNM) model \cite{hodge} and the Kovacs-Aklonis-Hutchinson-Ramos (KAHR) model \cite{kahr}. The Gutzow-Schmelzer approach \cite{gutzow}, theoretically better founded, replaces the fictive temperature by an order parameter. All three approaches characterize the glass by one single quantity, a fictive temperature, a deviation from the equilibrium density or an order parameter.

Very recently, a new physical picture for the highly viscous flow, the irreversible Eshelby model, has been developed  \cite{asyth,asyth1} and applied to the fall out of equilibrium \cite{asta}. The picture is based on Eshelby's \cite{eshelby} treatment of a structural rearrangement which changes the elastic misfit of the rearranging region to the elastic surroundings. Using the irreversible Eshelby model \cite{asyth,asyth1}, it is possible to describe the fall out of equilibrium with only two Vogel-Fulcher parameters \cite{asta}.

The present paper is an extension of this theoretical description \cite{asta}, replacing the single average fictive temperature by different fictive temperatures for different frozen Eshelby domains. It will be shown that one needs this extension for a successful description of the heating of a frozen glass back into equilibrium.

To describe non-equilibrium cooling data, one considers the Eshelby ensemble and its average fictive temperature $T_f$ for a given cooling rate $q=\partial T/\partial t$ in the neighborhood of the temperature $T_g(q)$, where the excess heat capacity $\Delta c_p$ of the undercooled liquid over the phonon heat capacity of the glass reaches half of its thermal equilibrium value. At this temperature, $\dot{T_f}=q/2$.

The time development of the average fictive temperature follows from the differential equation
\begin{equation} \label{diff}
	\dot{T_f}=\frac{T-T_f}{\tau_c(T_f,T)}
\end{equation}
because $\tau_c$ was defined in reference \cite{asyth} as the inverse of the average decay rate.

An additional basis of the treatment is the empirical Vogel-Fulcher law
\begin{equation}\label{vf}
	\ln{\tau_c}=\frac{B}{T-T_{VF}}-13\ln{10}
\end{equation}
corresponding to the Arrhenius barrier
\begin{equation}\label{vct}
   V_c(T)=Bk_BT/(T-T_{VF}).	
\end{equation}

The relaxation time $\tau_c(T_f,T)$ of the Eshelby ensemble is given by the Arrhenius relation
\begin{equation}\label{tctft}
\tau_c(T_f,T)=\tau_0\exp(V_c(T_f,T)/k_BT).	
\end{equation}
If $T$ is still closely below $T_f$, the energy barrier $V_c(T_f,T)$ depends only on $T_f$ and can be determined from eq. (\ref{vct}), and $\tau_0=10^{-13}$ s. 

But this simple scheme fails when $T$ leaves the neighborhood of $T_f$. As will be shown in the following, it is necessary to postulate
\begin{equation}\label{vctf}
	V_c(T_f,T)=bV_c(T_f)+(1-b)V_c(T),
\end{equation}
with $b=2/3$, a barrier composed to two thirds by the barrier at $T_f$ and to one third of the barrier at the phonon temperature $T$. Eq. (\ref{vctf}) returns to eq. (\ref{vct}) when the two temperatures are close together.

With this $V_c(T_f,T)$, eq. (\ref{tctft}) provides the $\tau_c(T_f,T)$ which one has to insert into eq. (\ref{diff}) to calculate the time dependence of the average fictive temperature at a constant cooling rate $q=\dot{T}$, beginning at high temperature where the difference between $T_f$ and $T$ is negligible. The differential equation is solved numerically for small temperature steps, integrating the exponential decay over the corresponding time step with the average $\tau_c$ between the two temperatures.

As shown in the preceding paper \cite{asta}, this simple procedure is able to reproduce measured non-equilibrium $T_g(q)$-data from cooling scans, having nothing else than the two Vogel-Fulcher parameters determined from equilibrium TMDSC-scans. This was demonstrated for the first methodical comparison of equilibrium and cooling measurements, done by Hensel and Schick \cite{hensel} for five very different glass formers, two polymers, two silicate glasses and one ionic glass former.

What does not work, however, is the successful calculation of heating scans. It is straightforward to calculate a heating scan from eq. (\ref{diff}), starting with the fictive temperature obtained by cooling with the rate $q$. Fig. 1 compares polystyrene cooling and reheating $c_p^{red}$-data \cite{tropin} ($c_p^{red}$-data means data from which the glass heat capacity has been subtracted and which have been normalized to the difference $\Delta c_p$ between liquid and glass) to an irreversible Eshelby model calculation for $B$=1837 K and $T_{VF}$=320.5 K. The cooling curve is well described with these two parameters, but the heating curve calculated for a single average fictive temperature (starting the numerical calculation at the fictive temperature) has a peak which is much sharper than the experimental one.

\begin{figure}   
\hspace{-0cm} \vspace{0cm} \epsfig{file=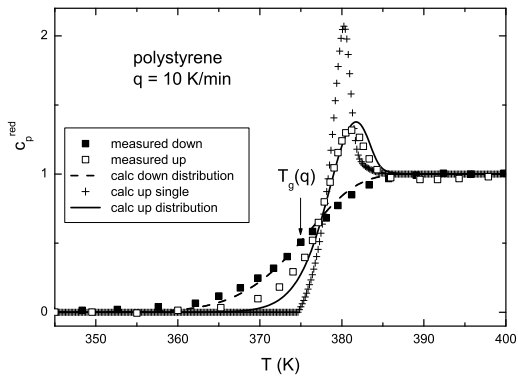,width=7 cm,angle=0} \vspace{0cm} \caption{Comparison of the two calculations for a single fictive temperature and a whole distribution, respectively, with each other and with measured data \cite{tropin}. The single fictive temperature concept (crosses) is able to describe the cooling data, but fails to describe the heating data.}
\end{figure}

In order to calculate the appropriate distribution of fictive temperatures, remember that the irreversible Eshelby transitions have a whole distribution of relaxation times around $\tau_c$
\begin{equation}\label{pt}
p(\ln{\tau})=\frac{\tau^2}{3\sqrt{2\pi}\tau_c^2}\left(\ln{\frac{4\sqrt{2}\tau_c}{\tau}}\right)^{3/2}
\end{equation}
(eq. (16) of reference \cite{asyth}), and that the average Eshelby domain at the rate $1/\tau_c$ freezes at $T_g(q)$ with 
\begin{equation}
	\dot{\tau}_c=1.54
\end{equation}
(eq. (12) of reference \cite{asta}). It follows that an Eshelby domain with the lifetime $\tau$ freezes at the temperature where
\begin{equation}
	\dot{\tau}_c=1.54\frac{\tau_c}{\tau},
\end{equation}
because $\dot{\tau}/\dot{\tau}_c=\tau/\tau_c$.

The numerical cooling calculation for the average fictive temperature supplies $\dot{\tau}_c$-values for all times. Thus one can tell at which time the domain freezes, and what the difference between phonon temperature $T$ and average fictive temperature $T_f$ is at this time.

With this knowledge, one can calculate for a given time interval $dt$, corresponding in phonon temperature to a temperature interval $dT$, the interval $d\ln{\tau}$ of the decaying Eshelby domains. The corresponding $p(\ln{\tau})d\ln{\tau}$ has to be multiplied with $T_f-T$ to get the resulting fictive temperature change, because, on the average, the domain starts at the average fictive temperature and lands at the phonon temperature.

\begin{figure}   
\hspace{-0cm} \vspace{0cm} \epsfig{file=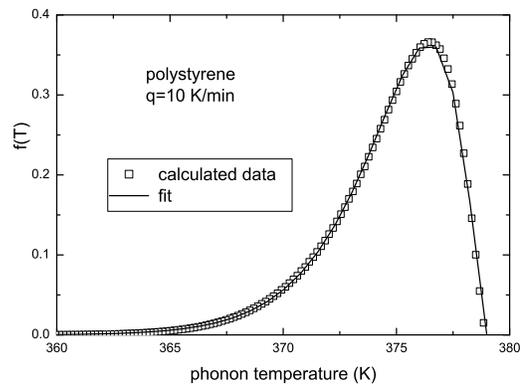,width=7 cm,angle=0} \vspace{0cm} \caption{Calculated function $f(T)$ for $q$ = 10 K/min in polystyrene and its fit in terms of eq. (\ref{ft}), with $T_g=374.9$ K and $\Delta T$=1.7 K.}
\end{figure}

Fig. 2 shows the result of this calculation, which supplies the distribution of the freezing signal over the phonon temperatures on cooling. It is well described by the fit function
\begin{equation}\label{ft}
	f(T)=f_0\exp((T_g-T)/\Delta T)(T_g+2.4\Delta T-T)^{3/2},
\end{equation}
where $T_g$ is $T_g(q)$ and $\Delta T$ decreases with decreasing cooling rate. If one folds this function with the cooling curve calculated for a single average fictive temperature, the curve broadens a bit, but leaves $T_g(q)$ unchanged. For the folding procedure, only $\Delta T$ is needed.

If one takes $f(T)$ as the distribution of fictive temperatures generated in the cooling process, it seems appropriate to fold $f(T)$ also with the curve obtained for the heating of the average fictive temperature. In fact, as seen in Fig. 1, the result lies much closer to the experimental data than the unfolded curve. 

\begin{figure}   
\hspace{-0cm} \vspace{0cm} \epsfig{file=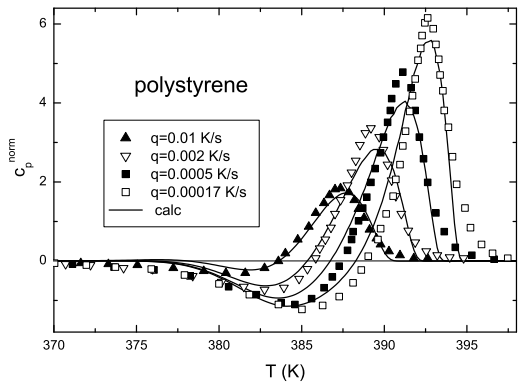,width=7 cm,angle=0} \vspace{0cm} \caption{Giant peaks \cite{tropin} in $c_p^{norm}$ (definition see text) and their description in terms of the Irreversible Eshelby mechanism.}
\end{figure}

The same procedure works also for a much more demanding case, the giant peaks in $c_p^{norm}$ in Fig. 3 of reference \cite{tropin}. $c_p^{norm}$ is again normalized to the difference $\Delta c_p$ between liquid and glass. It is obtained by first cooling with the slow cooling rate $q$, and then heating with 0.5 K/s, faster than $q$. Afterward, the sample is cooled with 0.5 K/s, and again heated with 0.5 K/s. The signal of the second heating run is subtracted from the signal of the first heating run, which implies that for $q$= 0.5 K/s $c_p^{norm}$ should be zero.

To calculate what the result should be, one first calculates cooling and heating with 0.5 K/s. $f(T)$ is determined from the cooling calculation and folded with the heating curve. This provides the data for the subtraction.

In the second step, one calculates the cooling scan with $q$ and obtains the $f_q(T)$ for this $q$. One finds that $\Delta T$ decreases from 1.45 K at $q=0.01$ K/s to 1.18 K at $q=0.00017$ K/s. Then, starting from the average $T_f$ at the end of the cooling scan, one calculates the heating scan for the average fictive temperature with 0.5 K/s, folds the result with $f_q(T)$ and subtracts the subtraction data. The result is compared to the data in Fig. 3.

The agreement is not perfect, but remember that there is only one single fitting parameter, namely the $b$ in eq. (\ref{vctf}). In fact, the giant peaks, which stretch over a large temperature range, fix $b$ with an accuracy of one percent.

To conclude, the ability of the irreversible Eshelby mechanism to describe out-of-equilibrium situations has been extended from cooling data to heating data.

\end{document}